\documentclass[copyright,creativecommons]{eptcs}
\providecommand{\event}{QFM 2009}
\providecommand{\volume}{?}
\providecommand{\anno}{2009}
\providecommand{\firstpage}{1}

\usepackage{amsmath}
\usepackage{amsxtra}
\usepackage{amstext}
\usepackage{amssymb}
\usepackage{latexsym}
\usepackage{setspace}
\usepackage[dvips]{graphicx}
\usepackage{breakurl}

\title { Verifying Real-Time Systems using Explicit-time Description Methods }
\author{ Hao Wang and Wendy MacCaull
         \institute{Centre for Logic and Information\\
         St. Francis Xavier University\\
         Antigonish, Canada }
         \email{\{hwang, wmaccaul\}@stfx.ca}
}

\def\titlerunning{Verifying Real-Time Systems using Explicit-time Description Methods}
\def\authorrunning{H. Wang \& W. MacCaull}
\begin{document}
\maketitle

\begin{abstract}
Timed model checking has been extensively researched in recent years. Many new formalisms with time extensions and tools based on them have been presented. On the other hand, \emph{Explicit-Time Description Methods} aim to verify real-time systems with general untimed model checkers. Lamport presented an explicit-time description method using a clock-ticking process (\emph{Tick}) to simulate the passage of time together with a group of global variables for time requirements. This paper proposes a new explicit-time description method with no reliance on global variables. Instead, it uses rendezvous synchronization steps between the \emph{Tick} process and each system process to simulate time. This new method achieves better \emph{modularity} and facilitates usage of more complex timing constraints. The two explicit-time description methods are implemented in {\sc DiVinE}, a well-known distributed-memory model checker. Preliminary experiment results show that our new method, with better modularity, is comparable to Lamport's method with respect to time and memory efficiency.
\end{abstract}

\section{Introduction}\label{SEC:introduction}

Model checking is an automatic analysis method which explores all possible states of a modeled system to verify whether the system satisfies a formally specified property. It was popularized in industrial applications, e.g., for computer hardware and software, and has great potential for modeling and monitoring complex and distributed business processes. \emph{Timed} model checking, the method to formally verify real-time systems, is attracting increasing attention from both the model checking community and the real-time community. However, general model checkers like SPIN \cite{Holzmann91BKspin} can only represent and verify the \emph{qualitative} relations between events, which constrains their use for real-time systems. The \emph{quantified} time notions, including time instant and duration, must be taken into account for timed model checking. For example in a safety critical application such as in an emergency department, after an emergency case arrives at the hospital, general model checking of hospital protocol can only verify whether ``the patient receives a certain treatment'', but to save the patient's life, it should be verified whether the protocol ensures that ``the patient receives a certain treatment within 1 hour''.

Many formalisms with time extensions have been presented as the basis for timed model checkers. A typical example is \emph{timed automata} \cite{DBLP:journals/tcs/AlurD94}, which is an extension of finite-state automata with a set of clock variables to keep track of time. Lamport \cite{Lamport05TRrealSimple} calls this approach as \emph{Implicit-Time Description Methods}. UPPAAL \cite{bllpwDimacs95uppaal} is a well-known timed-automata-based model checker; it has been successfully applied to various real-time controllers and communication protocols. Conventional temporal logics like \emph{Linear Temporal Logic} (LTL) or \emph{Computation Tree Logic} (CTL) must be extended \cite{DBLP:conf/rex/AlurH91} to handle the specification of properties of timed automata. The foundation for the decidability results in timed automata is based on the notion of \emph{region equivalence} over clock assignment \cite{BengtssonY03timedAutomata}. Models in a timed-automata-based model checker can not represent which time instant a transition is executed at within a time region; such model checkers can only deal with specification involving a time region or a pre-specified time instant. However, many real-time systems, especially those with pre-emptive scheduling features, need to record the time instant when the pre-emption happens for succeeding calculation. For example, triage is widely practiced in medical procedures; the caregiver \emph{C} may be administering some required but non-critical treatment on patient \emph{A} when another patient \emph{B} presents with a critical situation, such as a cardiac arrest. \emph{C} then must move to the higher priority task of treating \emph{B}, but it is necessary to store the elapsed time of \emph{A}'s treatment to determine how much time is still needed or the treatment needs to be restarted. The \emph{stop-watch} automata \cite{DBLP:conf/tacas/AbdeddaimM02StopWatchA}, an extension of timed autamata, is proposed to tackle this; unfortunately as Krc{\'a}l and Yi discussed in \cite{DBLP:conf/tacas/KrcalY04decidableTA}, since the reachability problem for this class of automata is undecidable, there is no guarantee for termination in the general case.

On the other hand, Lamport \cite{Lamport05TRrealSimple} advocated the Explicit-Time Description Methods which aim to use ordinary model checkers to realize timed model checking. He presented an explicit-time description method using a clock-ticking process (\emph{Tick}) to simulate the passage of time and a pair of global variables to store the time lower and upper bounds for each modeled system process. The main advantage of the explicit-time approach is that it does \emph{not} need specialized languages or tools for time description. The method has been implemented with popular model checkers SPIN (sequential) \cite{Holzmann91BKspin} and SMV \cite{McMillan92THsmv}. Recently, Van den Berg et al. \cite{DBLP:conf/fmics/BergSW07LEDMcaseStudy} successfully applied LEDM to verify the safety of railway interlockings for one of Australia's largest railway companies. The additional benefit of the explicit-time approach is that as it explicitly records the passage of time so the current time instant can be accessed easily, the pre-emptive scheduling problem discussed in the previous paragraph that causes difficulty using the timed-automata-based model checkers can be modeled naturally with explicit-time description methods.

In this paper, we propose a new explicit-time description method called \emph{Sync-based Explicit-time Description Method} (SEDM), which does not rely on global variables; instead it uses rendezvous synchronization steps between the \emph{Tick} process and each system process. After the \emph{Tick} process completes synchronization steps with every system processes, the global clock increments by one time unit. While, as Lamport commented \cite{Lamport05TRrealSimple}, ``{\it The approach (LEDM) cannot be used in process-based languages and formalisms with no explicit global state, such as CCS, CSP, Petri nets, streams and I/O automata}'', SEDM can do exactly that. As an added advantage, SEDM allows the timing constraints to be defined either globally or locally so the whole system can be modeled in a way that enhances its modularity. We choose {\sc DiVinE} \cite{Barnat2006Divine}, a well-known distributed model checker, because it accommodates the up-to-date multi-core architecture, i.e., clusters of multi-core CPU's and it has been tested successfully in large-scale clusters, even in a large-scale optical grid \cite{VBBBipdps09divine}. Experimental results show that SEDM is comparable to LEDM with respect to time and memory efficiency so SEDM can be used in place of LEDM.

The remainder of the paper is organized as follows. After a brief introduction to {\sc DiVinE}, Section 2 presents the LEDM with its {\sc DiVinE} implementation. The new method SEDM with its {\sc DiVinE} implementation is presented in Section 3. Section 4 describes our experiments and the results. Section 5 concludes the paper.

\section{Preliminaries}\label{SEC:preliminary}

The syntax outlined in \ref{SUBSEC:preDVE}, being incomplete, is meant for the presentation of the time-explicit description methods; the complete description can be found in \cite{divinePrjPage}.

\subsection{The {\sc DiVinE} Model Checker and its Modeling Language}\label{SUBSEC:preDVE}

{\sc DiVinE} is an explicit-state LTL model checkers based on the automata-based procedure by Vardi and Wolper \cite{DBLP:conf/lics/VardiW86}. The property to be specified is described by an LTL formula, both the system model and the LTL formula are represented by automata, then the model checking problem is reduced to detecting in the combined automaton graph whether there is an \emph{accepting cycle}, i.e., a cycle in which one of the vertices is marked ``accepting''. With the distributed algorithms to assign different portions of the state space to be explored by different machines, {\sc DiVinE} can: (1) verify much larger system models; (2) finish the verification in significantly less time (in comparison with the well-known explicit-state LTL model checker SPIN).

DVE is the modeling language of {\sc DiVinE}. Like in Promela (the modeling language of SPIN), a model described in DVE consists of processes, message channels and variables. Each process, identified by a unique name $procid$, consists of a list of local variable declarations, process states declarations, initial state declaration and a list of transitions. A transition transfers the process state from $ stateid_{1}$ to $ stateid_{2}$, the transition may contain a guard (which decides whether the transition can be executed), a synchronization (which communicates data with another process) and effects (which assigns new values to local or global variables). So we have

\

{\tt \ Transition ::= $ stateid_{1}$  -> $ stateid_{2}$  \{ Guard Sync Effect \} }

\

The {\tt Guard} contains the keyword {\tt guard} followed by a boolean expression and the {\tt Effect} contains the keyword {\tt effect} followed by a list of assignments. The {\tt Sync} follows the denotation for communication in CSP, `!' for the sender and `?' for the receiver. The synchronization can be either asynchronous or rendezvous. The $chanid$ is the channel for the synchronization; value(s) can be transferred in it. So we have

\

{\tt \ Sync ::= sync $chanid$!SyncValue $\vert$ $chanid$?SyncValue }

\

The property to be specified can be written as an LTL formula and a corresponding \emph{property process} can be automatically generated. Modeled system processes and the property process progress synchronously, so the latter can observe the system's behavior step by step and catch errors.

\subsection{Lamport Explicit-time Description Method}\label{SUBSEC:timeLamport}

The passage of time and timed quantified values can be expressed in untimed languages and properties to be specified can be expressed in conventional temporal logics. In LEDM, current time is represented with a global variable \emph{now} that is incremented by an added \emph{Tick} process. As we mentioned earlier, ordinary model checkers can only deal with integer variables, and the real-time system can be modeled in discrete-time only using an explicit-time description. The \emph{Tick} process increments \emph{now} by 1.

\begin{figure}
\begin{center}
  \includegraphics[width=3in]{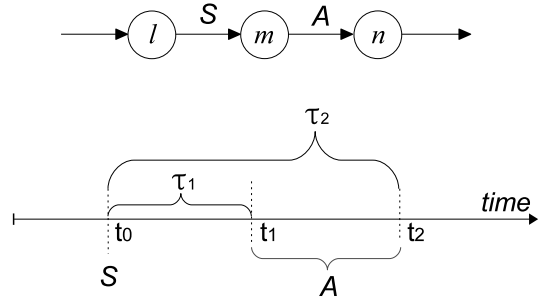}\\
  \caption{States and timeline for process $P_{i}$}
  \label{Fig:timelineLamport}
\end{center}
\end{figure}

Placing lower-bound and upper-bound timing constraints on transitions in processes is the common way to model real-time systems. Figure \ref{Fig:timelineLamport} shows a simple example of only two transitions, transition \emph{S}: {\tt $ stateid_{l}$  -> $ stateid_{m}$} is followed by the transition \emph{A}: {\tt $ stateid_{m}$  -> $ stateid_{n}$}. An upper-bound timing constraint on when a transition \emph{A}: {\tt $ stateid_{m}$  -> $ stateid_{n}$} must occur is expressed by a guard on the transition in the \emph{Tick} process so as to prevent an increase in time from violating the constraint. A lower-bound constraint on when the transition \emph{A} may occur is expressed by a guard on \emph{A} so it cannot be executed earlier than it should be. Each system process $P_{i}$ has a pair of count-down timers as global variables $ubtimer_{i}$ and $lbtimer_{i}$ for the timing constraints on its transitions. A large enough integer constant {\tt INFINITY} is defined; those upper bound timers with the value of {\tt INFINITY} are not active and the \emph{Tick} process does not decrement them. All upper bound timers are initialized to {\tt INFINITY} and all lower bound timers are initialized to zero. For transition \emph{A}, the timers will be set to the correct values by its preceding transition \emph{S}. As \emph{now} is incremented by 1, each non{\tt -INFINITY} {\tt ubtimer} and non-zero {\tt lbtimer} is decremented by 1.

Initially, $(ubtimer_{i},lbtimer_{i})$ are set to $({\tt INFINITY},0)$. The transition \emph{S} is executed at time instant $t_{0}$, and $(ubtimer_{i},lbtimer_{i})$ are set to $(\tau_{2},\tau_{1})$. After $\tau_{1}$ time units, i.e., at time instant $t_{1}$ when $(ubtimer_{i},lbtimer_{i})$ is equal to $(\tau_{2}-\tau_{1},0)$, the transition \emph{A} is enabled. Both timers will be reset or set to new time bounds after the execution of \emph{A}. If the transition \emph{A} is still not executed when the time reaches $t_{2}$ and $ubtimer_{i}$ is equal to 0, the transition in the \emph{Tick} process is disabled, which means the clock has to stop here. Only after $ubtimer_{i}$ is set by transition \emph{A}, the \emph{Tick} process can start again. In this way, the time upper-bound constraint is realized.

The \emph{Tick} process and the system process $P_{i}$ in DVE are described in Figure \ref{Fig:tickProcLamportMethod} and Figure \ref{Fig:sysProcLamportMethod}.

\begin{figure}[htp]
{\tt \ process P\_Tick \{ }

{\tt \ \ \ state tick; }

{\tt \ \ \ init tick; }

{\tt \ \ \ trans }

{\tt \ \ \ \ \ tick -> tick \{ guard $all$  $ubtimers$ >0;}

{\tt \hspace{110 pt} effect now = now + 1, }

{\tt \hspace{125 pt} $decrements$ $all$ $timers$; \} ; }

{\tt \ \}  }
\caption[]{\emph{Tick} process in DVE for LEDM \label{Fig:tickProcLamportMethod}}
\end{figure}

\begin{figure}[htp]
{\tt \ process P\_i \{ }

{\tt \ \ \ state ..., state\_l, state\_m, state\_n; }

{\tt \ \ \ init ...; }

{\tt \ \ \ trans }

{\tt \hspace{38 pt} ... -> ... ; }

{\tt \ \ \ \ \ state\_l -> state\_m \{ ...; effect $set$ $timers$ $for$ $transitionA$;\}, }

{\tt \ \ \ \ \ state\_m -> state\_n \{ guard lbtimer$[i]$==0; effect ... ; \},}

{\tt \hspace{38 pt} ... -> ... ; }

{\tt \ \}  }
\caption[]{System process $P_{i}$ in DVE for LEDM \label{Fig:sysProcLamportMethod}}
\end{figure}

We observe that the value of \emph{now} is limited by the size of type {\tt integer} and careless incrementing can cause overflow error. This can be avoided by incrementing \emph{now} using modular arithmetic, i.e., setting $now = (now+1)$ {\tt mod MAXIMAL} ({\tt MAXIMAL} is the maximal integer value supported by the model checker). The value limit can also be increased by linking several integers, i.e., every time {\tt ($int_1$+1) mod MAXIMAL} becomes zero again,  $int_2$ increments by 1, and so on. Note that the variable \emph{now} is only incremented in the \emph{Tick} process and does not appear in any other process. So for general system models in which time lower and upper bounds suffice, the variable \emph{now} should be removed.

\section{The New Sync-based Explicit-Time Description Method}\label{SEC:SEDM}

This section presents the new SEDM, followed by two examples to illustrate its modularity advantage and capability to model pre-emptive scheduling problems.

\subsection{The Method}\label{SUBSEC:method}

In the new SEDM, the passage of time is also simulated by an additional \emph{Tick} process. In one time unit, it completes synchronization steps with each system process. The current time is the count of previous synchronization steps, so all the timing variables can be defined either locally or globally. In this way, local timers can be added or removed without affecting the model globally and good \emph{modularity} can be achieved. Note that the \emph{now} variable can also be removed for a similar reason, but if any system process contains any enabling condition that is dependent on a certain time instant, it is safe to define a \emph{now} variable locally.

For the same example in Figure \ref{Fig:timelineLamport}, $P_{i}$ has local timers $(ubtimer,lbtimer)$. For the transition \emph{A}: {\tt $ stateid_{m}$  -> $ stateid_{n}$}, each of the timers will be set to the correct values $(\tau_{2},\tau_{1})$ by its preceding transition, \emph{S}: {\tt $ stateid_{l}$  -> $ stateid_{m}$}. The execution is similar to Lamport's method except: (1) the timers are decremented locally by 1 after each synchronization with the \emph{Tick} process; (2) if the transition \emph{A} is still not executed when the time reaches $t_{2}$ and $ubtimer_{i}$ is equal to 0, there is no synchronization step before executing transition \emph{A}. Because the \emph{Tick} process has to synchronize with each process for each tick, it must wait for $P_{i}$'s next {\tt sync} statement.

The \emph{Tick} process, for two system processes, in DVE is described in Figure \ref{Fig:tickProcNewMethod}. The local {\tt ubtimer} and {\tt lbtimer} can be defined and used in a system process as in Figure \ref{Fig:systemProcNewMethod}.

\begin{figure}[h!]
{\tt \ process P\_Tick \{ }

{\tt \ \ \ state tick1, tick2; }

{\tt \ \ \ init tick1; }

{\tt \ \ \ trans }

{\tt \ \ \ \ \ tick1 -> tick2 \{ sync chan1!; \}, }

{\tt \ \ \ \ \ tick2 -> tick1 \{ sync chan2!; \}; }

{\tt \ \}  }
\caption[]{\emph{Tick} process in DVE for SEDM \label{Fig:tickProcNewMethod}}
\end{figure}

\begin{figure}[h!]
{\tt \ process P\_i \{ }

{\tt \ \ \ int ubtimer, lbtimer; }

{\tt \ \ \ state state\_l, state\_m, state\_n, ...; }

{\tt \ \ \ init ...; }

{\tt \ \ \ trans }

{\tt \hspace{34 pt} ... \  -> \ ... ; }

{\tt \ \ \ \ \ state\_l -> state\_m \{ ...; effect $set$ $timers$ $for$ $transitionA$ ; \},}

{\tt \ \ \ \ \ state\_m -> state\_m \{ guard ubtimer>0; sync chan1? ;}

{\tt \hspace{140 pt} effect $decrement$ $timers$ $by$ $1$ ; \}, }

{\tt \ \ \ \ \ state\_m -> state\_n \{ guard lbtimer==0 \&\& ...; ...; \},}

{\tt \hspace{34 pt} ... \  -> \ ... ; }

{\tt \ \}  }
\caption[]{System process $P_{i}$ in DVE for SEDM \label{Fig:systemProcNewMethod}}
\end{figure}

Readers may argue against the usage of round-robin scheduling of all synchronization steps in one tick: P\_1 always ticks before P\_2. Actually, a time model to be verified is built to cover every possible execution of all system steps, which can be assured in SEDM by separating transitions for system steps and transitions for time synchronization in all system processes. Therefore, we do not need to cover every possible sequence of all synchronization steps, one sequence is enough for the verification.

Readers may also be concerned about the size of the state space and time efficiency as SEDM adds \emph{N} synchronization steps for every time unit, \emph{N} being the number of system processes. However, the experimental results (see Section \ref{SEC:exprDesc}) show that as the model grows bigger, the time and memory efficiency and size of state space are comparable to those of LEDM.

\subsection{An Example with Complex Timers}\label{SUBSEC:complexTimers}

As the time can be accessed locally with SEDM, complex timing constraints, e.g., fixed time delay (the special case when {\tt ubtimer==lbtimer}), multiple independent (possibly overlapping) timers and dependent timers, can be expressed more conveniently than with LEDM because with the latter method new global variables must defined and the {\emph{Tick} process must be updated.

\begin{figure}[h!]
\begin{center}
  \includegraphics[width=3in]{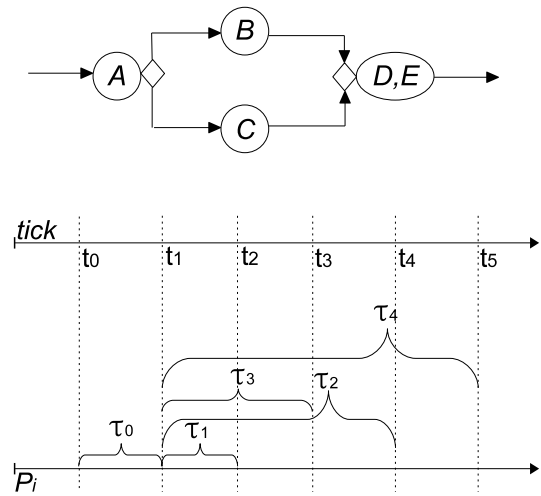}\\
  \caption{States and timeline for complex timers using SEDM}\label{Fig:timelineNewMethod2}
\end{center}
\end{figure}

\begin{figure}[h!]
{\tt \ process P\_i \{ }

{\tt \ \ \ ...; }

{\tt \ \ \ trans }

{\tt \hspace{34 pt} ... \  -> \ ... ; }

{\tt \ \ \ \ \ state\_l -> state\_m \{ ...; effect fixdelay=$\tau_{0}$; \},}

{\tt \ \ \ \ \ state\_m -> state\_m \{ guard fixdelay>0; sync chan1?; }

{\tt \hspace{140 pt} effect fixdelay=fixdelay-1 ; \},}

{\tt \ \ \ \ \ state\_m -> state\_n \{ guard fixdelay==0 ; ...; }

{\tt \hspace{140 pt} effect ubtimer1=$\tau_{3}$,lbtimer1=$\tau_{1}$,}

{\tt \hspace{160 pt} ubtimer2=$\tau_{4}$,lbtimer2=$\tau_{2}$; \}, }

{\tt \ \ \ \ \ state\_n -> state\_n \{ guard ubtimer2>0; sync chan1?; }

{\tt \hspace{140 pt} effect $decrement$ $timers$ $by$ $1$;\}, }

{\tt \ \ \ \ \ state\_n -> state\_o \{ guard ubtimer1>0 \&\& lbtimer1==0; ...;\},}

{\tt \ \ \ \ \ state\_n -> state\_p \{ guard ubtimer2>0 \&\& lbtimer2==0; ...;\},}

{\tt \hspace{34 pt} ... \  -> \ ... ; }

{\tt \ \}  }
\caption[]{System process $P_{i}$ in DVE with complex timers \label{Fig:systemProcNewMethod2}}
\end{figure}

Figure \ref{Fig:timelineNewMethod2} describes five transitions $A, B, C, D, E$ in $P_{i}$ (see the upper part of the figure) and their associated timeline. Transition $A$: {\tt $ stateid_{m}$  -> $ stateid_{n}$} has a fixed time delay, $\tau_{0}$; transition $B$: {\tt $ stateid_{n}$  -> $ stateid_{o}$} has upper and lower bounds, $(\tau_{2},\tau_{1})$; transition $C$: {\tt $ stateid_{n}$  -> $ stateid_{p}$} has upper and lower bounds, $(\tau_{4},\tau_{3})$. After the execution of transition $A$, there is a time period, $(t_{3},t_{4})$, during which both transition $B$ and $C$ are enabled and chosen non-deterministically. Transition $D$: {\tt $ stateid_{o}$  -> $ stateid_{q}$} and $E$: {\tt $ stateid_{p}$  -> $ stateid_{q}$} have the upper and lower bounds which are dependant on the execution time of $B$ or $C$. The process $P_{i}$ in DVE is described in Figure \ref{Fig:systemProcNewMethod2}.

\subsection{An Example of Pre-emptive Scheduling}\label{SUBSEC:preemptiveExample}

Following the triage example described in Section \ref{SEC:introduction}, we consider a system of multiple parallel tasks with different priorities, assuming that the right to an exclusive resource is deprivable, i.e., a higher priority task \emph{B} may deprive the resource from the currently running task \emph{A}. In this case, the elapsed time of \emph{A}'s execution must be stored for a future resumed execution.

\begin{figure}[h!]
\begin{center}
  \includegraphics[width=3in]{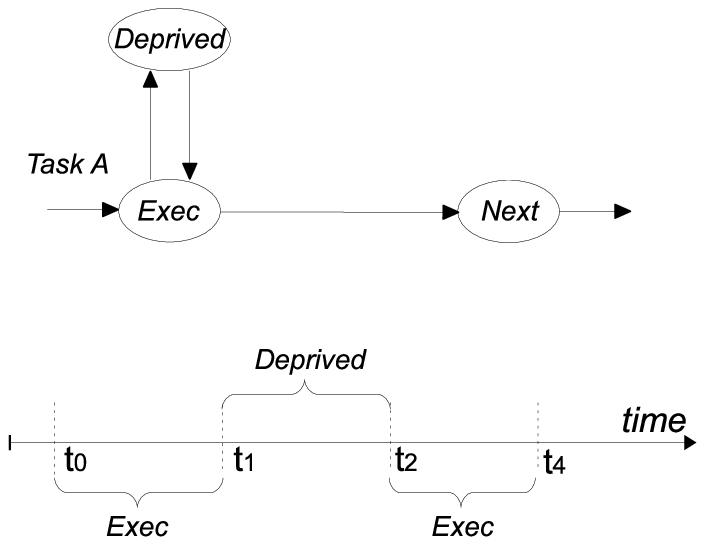}\\
  \caption{An Example of Pre-emptive Scheduling}\label{Fig:deprivableScheduling}
\end{center}
\end{figure}

\begin{figure}[h!]
{\tt \hspace{16 pt} byte isROccupied=0; $//$0 means available}

{\tt \hspace{16 pt} process A \{ }

{\tt \hspace{30 pt} default(Tag,$tag_{A}$) }

{\tt \hspace{30 pt} int timeToGo=10; }

{\tt \hspace{30 pt} state s\_i, s\_Exec, s\_Deprived, ...; }

{\tt \hspace{30 pt} init ...; }

{\tt \hspace{30 pt} trans }

{\tt \hspace{50 pt} ... -> ... ; }

{\tt \hspace{50 pt} s\_i -> s\_Exec \{ guard isROccupied==0; }

{\tt \hspace{140 pt} effect isROccupied=Tag, ltimer=timeToGo; }

{\tt \hspace{50 pt} s\_Exec -> s\_Exec \{ guard ltimer>0; sync chan1?; }

{\tt \hspace{155 pt} effect ltimer=ltimer-1; \},}

{\tt \hspace{50 pt} s\_Exec -> s\_Deprived \{ guard isROccupied\!=Tag \&\& ltimer>0; }

{\tt \hspace{178 pt} effect timeToGO=ltimer; \}, }

{\tt \hspace{50 pt} s\_Deprived -> s\_Deprived \{ guard isROccupied!=0; sync chan1?; \}}

{\tt \hspace{50 pt} s\_Deprived -> s\_Exec \{ guard isROccupied==0; }

{\tt \hspace{178 pt} effect isROccupied=Tag, ltimer=timeToGo; \}, }

{\tt \hspace{50 pt} s\_Exec -> s\_Next \{ guard ltimer==0; }

{\tt \hspace{155 pt} effect isROccupied=0; \}, }

{\tt \hspace{50 pt} ... -> ... ; }

{\tt \hspace{16 pt} \}  }
\caption[]{Process in DVE for Pre-emptive Scheduling Example using SEDM \label{Fig:systemProcNewMethod3}}
\end{figure}

Figure \ref{Fig:deprivableScheduling} shows a portion of a state transition diagram for task \emph{A}, assuming \emph{A} needs the exclusive resource \emph{R} for 10 time units; when \emph{R} becomes available at time instant $t_{0}$, \emph{A} starts its execution by entering the state \emph{Exec}; at time instant $t_{1}$, \emph{B} deprives \emph{A}'s right to \emph{R}, and \emph{A} changes to the state \emph{Deprived} and stores the elapsed $t_{1}-t_{0}$ time units; when \emph{R} becomes available again, \emph{A} resumes its execution to state \emph{Exec} for the remaining $10-(t_{1}-t_{0})$ units. Implementation of this example using any one of the three explicit-time description methods is straightforward. Figure \ref{Fig:systemProcNewMethod3} shows the process for task \emph{A} in DVE using SEDM (assuming \emph{A} has the lowest priority).

\section{Experiments in {\sc DiVinE}}\label{SEC:exprDesc}

For the convenience of comparison, we experiment with the Fischer's mutual exclusion algorithm, a well-known benchmark for timed model checking, which is also used by Lamport in his experiments \cite{Lamport05TRrealSimple}. The brief description of the algorithm is adapted from \cite{Lamport05TRrealSimple}. Our experiments model the algorithm in {\sc DiVinE} using LEDM and SEDM, and compare the time and memory efficiency and size of state space.

Fischer's algorithm is a shared-memory, multi-threaded algorithm. It uses a shared variable \emph{x} whose value is either a thread identifier (starting from 1) or zero; its initial value is zero. For the convenience of specification of the safety property in our experiments, we use a counter \emph{c} to count the number of threads that are in the critical section.  The program for thread \emph{t} is described in Figure \ref{Fig:fischerAlgo}.

\begin{figure}[h!]
{ {\it \hspace{135 pt} ncs}: noncritical section;

  {\it \hspace{135 pt} a}: {\bf wait until} {\it x} = 0;

  {\it \hspace{135 pt} b}: {\it x} := {\it t};

  {\it \hspace{135 pt} c}: {\bf if} {\it x} $\neq$ {\it t} {\bf then goto} {\it a};

  {\it \hspace{135 pt} cs}: critical section;

  {\it \hspace{135 pt} d}: {\it x} := 0; {\bf goto} {\it ncs};
}
\caption[]{Program of thread \emph{t} in Fischer's algorithm \label{Fig:fischerAlgo}}
\end{figure}

The timing constraints are, first, that step \emph{b} must be executed at most $\delta$ time units (as a upper bound) after the preceding execution of step \emph{a}; and second, that step \emph{c} cannot be executed until at least $\epsilon$ time units (as a lower bound) after the preceding execution of step \emph{b}. For step \emph{c}, there is an additional upper bound $\epsilon_{upper}$ to ensure fairness. For convenience, we use the same value for three constraints, i.e., $\delta=\epsilon=\epsilon_{upper}=T$. The algorithm is tested for 6 threads. The safety property, ``no more than one process can be in the critical section'', is specified as $G (c<2)$ for the model.

\begin{figure*}[h!]
\begin{center}
\begin{tabular}{|r|r|r|r|r|r|r|}
  \hline
      & \multicolumn{3}{c|}{LEDM} & \multicolumn{3}{c|}{SEDM} \\ \cline{2-7}
  \emph{T} & States & {Time} & {Memory} & States & {Time} & {Memory} \\ \hline
  2 & 644987 & 1.8 & 4700.1 & 1838586 & 2.9 & 4865.3 \\
  4 & 3048515 & 3.3 & 4942.8 & 6923088 & 4.3 & 5641.9 \\
  6 & 11201179 & 7.2 & 6343.4 & 18460632 & 9.3 & 7402.0 \\
  8 & 32952899 & 18.6 & 9958.9 & 48177552 & 21.2 & 11905.0 \\
  10 & 82428155 & 49.2 & 18016.2 & 113914104 & 46.1 & 21894.8 \\
  12 & 182767747 & 115.0 & 34906.3 & 244265616 & 108.8 & 41454.5 \\
  14 & 369377435 & 290.9 & 65205.1 & 482259672 & 230.0 & 78936.2 \\
  16 & 693683459 & 617.5 & 122549.0 & 889586256 & 611.2 & 148010.0 \\
  \hline
\end{tabular}
\caption[]{Time (in seconds), number of states and memory usage (in MB) for Fischer's algorithm using two explicit-time methods in {\sc DiVinE} with 16 CPUs \label{Fig:timeResults2}}
\end{center}
\end{figure*}

The version 0.8.1 of the {\sc DiVinE}-Cluster is used. This version has the new feature of pre-compiling the model in DVE into dynamically linked C functions; this feature speeds up the state space generation significantly. According to the published experimental results of {\sc DiVinE} \cite{VBBBipdps09divine}, we choose the OWCTY ({\it One Way to Catch Them Young}) algorithm for better time efficiency as our example property is known to hold.

All experiments are executed on the Mahone cluster of ACEnet \cite{ACEnetPrjPage}, the high performance computing consortium for universities in Atlantic Canada. The cluster is a Parallel Sun x4100 AMD Opteron (dual-core) cluster equipped with Myri-10G interconnection. Parallel jobs are assigned using the Open MPI library.

Figure \ref{Fig:timeResults2} compares time and memory efficiency for the two explicit-time description methods in both versions of {\sc DiVinE} with 16 CPUs; it also shows how the size of state spaces increase as \emph{T} increases.

While SEDM has the bigger number of states for all models, as the model becomes larger, the time increases more slowly than with LEDM: time increases by a factor of 343 as \emph{T} increases from 2 to 16 with LEDM; time increases by a factor of 204 as \emph{T} increases from 2 to 16 with SEDM; It is also interesting to find that starting from $T=10$, the time spent with SEDM is \emph{less} than the time with LEDM.

Because SEDM adds \emph{N} synchronization steps (recall that \emph{N} is the number of system processes) for each time units, the size of state space of the model generated by our method is bigger than that by Lamport's method. But as the model becomes bigger, the difference becomes insignificant. For $T=2$, ${\tt states(SEDM)} \over {\tt states(LEDM)}$=2.85, while for $T=16$, the two numbers of state size become comparable.

The memory usages of both methods are comparable. Because OWCTY algorithm requires that the whole state space fit into the (distributed) memory, enough memory resource must be allocated in order for the verification to succeed.

Note that when increasing the number of CPUs an added portion of memory needs to be counted for increasing inter-node communications.

\section{Discussion and Conclusion}\label{SEC:conclude}

In this paper, we propose a new method, SEDM using rendezvous synchronization steps, so the timing constraints can be defined either globally or locally, compared to the heavy reliance on global variables in LEDM. Consequently, SEDM makes it possible to model discrete time with some process-based untimed languages without explicit global variables. With SEDM, real-time systems can be modeled with a high degree of modularity and more complex timing constraints can be modeled more conveniently.

As Lamport mention in \cite{Lamport05TRrealSimple}, the explicit-time description methods are not designed to beat specialized timed model checkers like UPPAAL: it is obvious that time-automata-based model checkers can handle continuous time semantics while EDMs can only deal with discrete time semantics. However, EDMs are intended to offer more options for the verification of real-time systems. First, explicit-time description methods provide a solution for accessing and storing the current time instant for the pre-emptive scheduling models. Second, while the size of state space in an explicit-time method grows along with the number of time units, it is less sensitive to the number of concurrently running timers. This suggests that the explicit-time method implemented in an un-timed model checker may verify more complex system behaviors. Third, as Van den Berg et al. mention in \cite{DBLP:conf/fmics/BergSW07LEDMcaseStudy}, in some real-world scenarios when significant resources already have been invested into the model for a general model checker such as SPIN or SMV, it is much easier and therefore preferable to extend the existing model to represent time notions rather than to re-model the entire system for a specialized timed model checker. Last but not least, explicit-time description methods enable the usage of existing large-scale distributed model checkers such as {\sc DiVinE} so that we can verify much bigger real-time systems.

This research is part of an ambitious research and development project, {\it Building Decision-support through Dynamic Workflow Systems for Health Care} \cite{dallien08initial}. Verification that the health care process design meets its specifications and monitoring the process to check specifications for each instance (patient) are essential. Real world health care workflow processes are highly dynamic and local changes are the norm. In addition to work in verification, members of our research group \cite{ourPrjPage} are currently investigating parallel and distributed approaches to reasoning about structured knowledge bases (ontologies). Interfacing these reasoners and distributed model checkers with workflow engines will permit runtime monitoring of complex, highly variable and safety critical processes. Currently, we are using explicit-time description methods to model and verify real-world health care processes.

As a continuous effort in practical timed model checking, we also study the efficiency problem of explicit-time descriptions and have made some progress based on optimizing the tick process \cite{Hao09EEDM}, so that EDMs can be applied to problems of larger scale. Dutertre and Sorea \cite{DBLP:DutertreS04caldrAutomata} and Clarke et al. \cite{clarke07abstraction} recently presented two different abstraction techniques for timed automata and the abstraction outcome can be verified using un-timed model checkers. We also intend to study the possibility of this kind of technique in distributed model checkers.

\section*{Acknowledgment}
This research is sponsored by NSERC, an Atlantic Computational Excellence Network (ACEnet) Post Doctoral Research Fellowship and by the Atlantic Canada Opportunities Agency through an Atlantic Innovation Fund project. The computational facilities are provided by ACEnet. We also thank Jiri Barnat, Keith Miller and the anonymous reviewers of QFM'09 for their helpful comments.

\bibliographystyle{eptcs}
\bibliography{all}

\begin{thebibliography}{10}
\providecommand{\bibitemstart}[1]{\bibitem{#1}}
\providecommand{\bibitemend}{}
\providecommand{\bibliographystart}{}
\providecommand{\bibliographyend}{}
\providecommand{\url}[1]{\texttt{#1}}
\providecommand{\urlprefix}{Available at }
\providecommand{\bibinfo}[2]{#2}
\bibliographystart

\bibitemstart{ACEnetPrjPage}
\emph{\bibinfo{title}{{Atlantic Computational Excellence network} ({ACEnet}).
  http://www.ace-net.ca/. {$Last$ $accessed$ $on$ $Nov.$ $2009$}.}}
\bibitemend

\bibitemstart{ourPrjPage}
\emph{\bibinfo{title}{{Centre for Logic and Information, St. Francis Xavier
  University.} http://logic.stfx.ca/. {$Last$ $accessed$ $on$ $Nov.$ $2009$}.}}
\bibitemend

\bibitemstart{divinePrjPage}
\emph{\bibinfo{title}{{\sc DiVinE} project. http://divine.fi.muni.cz/. {$Last$
  $accessed$ $on$ $Nov.$ $2009$}.}}
\bibitemend

\bibitemstart{DBLP:conf/tacas/AbdeddaimM02StopWatchA}
\bibinfo{author}{Yasmina Abdedda\"{\i}m} \& \bibinfo{author}{Oded Maler}
  (\bibinfo{year}{2002}): \emph{\bibinfo{title}{Preemptive Job-Shop Scheduling
  Using Stopwatch Automata}}.
\newblock In: \bibinfo{editor}{Joost-Pieter Katoen} \& \bibinfo{editor}{Perdita
  Stevens}, editors: {\sl \bibinfo{booktitle}{TACAS}}, {\sl
  \bibinfo{series}{Lecture Notes in Computer Science}} \bibinfo{volume}{2280}.
  \bibinfo{publisher}{Springer}, pp. \bibinfo{pages}{113--126}.
\bibitemend

\bibitemstart{DBLP:journals/tcs/AlurD94}
\bibinfo{author}{Rajeev Alur} \& \bibinfo{author}{David~L. Dill}
  (\bibinfo{year}{1994}): \emph{\bibinfo{title}{A Theory of Timed Automata}}.
\newblock {\sl \bibinfo{journal}{Theor. Comput. Sci.}}
  \bibinfo{volume}{126}(\bibinfo{number}{2}), pp. \bibinfo{pages}{183--235}.
\bibitemend

\bibitemstart{DBLP:conf/rex/AlurH91}
\bibinfo{author}{Rajeev Alur} \& \bibinfo{author}{Thomas~A. Henzinger}
  (\bibinfo{year}{1991}): \emph{\bibinfo{title}{Logics and Models of Real Time:
  A Survey}}.
\newblock In: \bibinfo{editor}{J.~W. de~Bakker}, \bibinfo{editor}{Cornelis
  Huizing}, \bibinfo{editor}{Willem~P. de~Roever} \& \bibinfo{editor}{Grzegorz
  Rozenberg}, editors: {\sl \bibinfo{booktitle}{REX Workshop}}, {\sl
  \bibinfo{series}{Lecture Notes in Computer Science}} \bibinfo{volume}{600}.
  \bibinfo{publisher}{Springer-Verlag}, pp. \bibinfo{pages}{74--106}.
\bibitemend

\bibitemstart{Barnat2006Divine}
\bibinfo{author}{Jiri Barnat}, \bibinfo{author}{Lubos Brim},
  \bibinfo{author}{Ivana \v{C}ern\'{a}}, \bibinfo{author}{Pavel Moravec},
  \bibinfo{author}{Petr Ro\v{c}kai} \& \bibinfo{author}{Pavel \v{S}ime\v{c}ek}
  (\bibinfo{year}{2006}): \emph{\bibinfo{title}{{DiVinE -- A Tool for
  Distributed Verification (Tool Paper)}}}.
\newblock In: {\sl \bibinfo{booktitle}{{Computer Aided Verification}}}, {\sl
  \bibinfo{series}{Lecture Notes in Computer Science}} \bibinfo{volume}{4144}.
  \bibinfo{publisher}{Springer-Verlag}, pp. \bibinfo{pages}{278--281}.
\bibitemend

\bibitemstart{bllpwDimacs95uppaal}
\bibinfo{author}{Johan Bengtsson}, \bibinfo{author}{Kim~G. Larsen},
  \bibinfo{author}{Fredrik Larsson}, \bibinfo{author}{Paul Pettersson} \&
  \bibinfo{author}{Wang Yi} (\bibinfo{year}{1995}): \emph{\bibinfo{title}{{{\sc
  Uppaal} --- a Tool Suite for Automatic Verification of Real--Time Systems}}}.
\newblock In: {\sl \bibinfo{booktitle}{Proc. of Workshop on Verification and
  Control of Hybrid Systems III}}, number \bibinfo{number}{1066} in
  \bibinfo{series}{Lecture Notes in Computer Science}.
  \bibinfo{publisher}{Springer-Verlag}, pp. \bibinfo{pages}{232--243}.
\bibitemend

\bibitemstart{BengtssonY03timedAutomata}
\bibinfo{author}{Johan Bengtsson} \& \bibinfo{author}{Wang Yi}
  (\bibinfo{year}{2003}): \emph{\bibinfo{title}{Timed Automata: Semantics,
  Algorithms and Tools}}.
\newblock In: \bibinfo{editor}{J{\"o}rg Desel}, \bibinfo{editor}{Wolfgang
  Reisig} \& \bibinfo{editor}{Grzegorz Rozenberg}, editors: {\sl
  \bibinfo{booktitle}{Lectures on Concurrency and Petri Nets}}, {\sl
  \bibinfo{series}{Lecture Notes in Computer Science}} \bibinfo{volume}{3098}.
  \bibinfo{publisher}{Springer}, pp. \bibinfo{pages}{87--124}.
\bibitemend

\bibitemstart{DBLP:conf/fmics/BergSW07LEDMcaseStudy}
\bibinfo{author}{Lionel van~den Berg}, \bibinfo{author}{Paul~A. Strooper} \&
  \bibinfo{author}{Kirsten Winter} (\bibinfo{year}{2007}):
  \emph{\bibinfo{title}{Introducing Time in an Industrial Application of
  Model-Checking}}.
\newblock In: \bibinfo{editor}{Stefan Leue} \& \bibinfo{editor}{Pedro Merino},
  editors: {\sl \bibinfo{booktitle}{FMICS}}, {\sl \bibinfo{series}{Lecture
  Notes in Computer Science}} \bibinfo{volume}{4916}.
  \bibinfo{publisher}{Springer}, pp. \bibinfo{pages}{56--67}.
\bibitemend

\bibitemstart{clarke07abstraction}
\bibinfo{author}{Edmund~M. Clarke}, \bibinfo{author}{Flavio Lerda} \&
  \bibinfo{author}{Muralidhar Talupur} (\bibinfo{year}{2007}):
  \emph{\bibinfo{title}{An Abstraction Technique for Real-time Verification}}.
\newblock In: \bibinfo{editor}{S.~Ramesh} \& \bibinfo{editor}{P.~Sampath},
  editors: {\sl \bibinfo{booktitle}{Next Generation Desigh and Verification
  Methodologies}}, Lecture Notes in Computer Science.
  \bibinfo{publisher}{Springer-Verlag}, pp. \bibinfo{pages}{1--17}.
\bibitemend

\bibitemstart{dallien08initial}
\bibinfo{author}{Jeff Dallien}, \bibinfo{author}{Wendy MacCaull} \&
  \bibinfo{author}{Allen Tien} (\bibinfo{year}{2008}):
  \emph{\bibinfo{title}{Initial Work in the Design and Development of
  Verifiable Workflow Management Systems and Some Applications to Health
  Care}}.
\newblock In: {\sl \bibinfo{booktitle}{5th International Workshop on
  Model-based Methodologies for Pervasive and Embedded Software}}.
  \bibinfo{publisher}{IEEE Computer Society}, pp. \bibinfo{pages}{78--91}.
\bibitemend

\bibitemstart{DBLP:DutertreS04caldrAutomata}
\bibinfo{author}{Bruno Dutertre} \& \bibinfo{author}{Maria Sorea}
  (\bibinfo{year}{2004}): \emph{\bibinfo{title}{Modeling and Verification of a
  Fault-Tolerant Real-Time Startup Protocol Using Calendar Automata}}.
\newblock In: \bibinfo{editor}{Yassine Lakhnech} \& \bibinfo{editor}{Sergio
  Yovine}, editors: {\sl \bibinfo{booktitle}{FORMATS/FTRTFT}}, {\sl
  \bibinfo{series}{Lecture Notes in Computer Science}} \bibinfo{volume}{3253}.
  \bibinfo{publisher}{Springer-Verlag}, pp. \bibinfo{pages}{199--214}.
\bibitemend

\bibitemstart{Holzmann91BKspin}
\bibinfo{author}{Gerard~J. Holzmann} (\bibinfo{year}{1991}):
  \emph{\bibinfo{title}{Design and Validation of Computer Protocols}}.
\newblock \bibinfo{publisher}{Prentice Hall}.
\bibitemend

\bibitemstart{DBLP:conf/tacas/KrcalY04decidableTA}
\bibinfo{author}{Pavel Krc{\'a}l} \& \bibinfo{author}{Wang Yi}
  (\bibinfo{year}{2004}): \emph{\bibinfo{title}{Decidable and Undecidable
  Problems in Schedulability Analysis Using Timed Automata}}.
\newblock In: \bibinfo{editor}{Kurt Jensen} \& \bibinfo{editor}{Andreas
  Podelski}, editors: {\sl \bibinfo{booktitle}{TACAS}}, {\sl
  \bibinfo{series}{Lecture Notes in Computer Science}} \bibinfo{volume}{2988}.
  \bibinfo{publisher}{Springer}, pp. \bibinfo{pages}{236--250}.
\bibitemend

\bibitemstart{Lamport05TRrealSimple}
\bibinfo{author}{Leslie Lamport} (\bibinfo{year}{2005}):
  \emph{\bibinfo{title}{Real-Time Model Checking is Really Simple}}.
\newblock In: \bibinfo{editor}{Dominique Borrione} \&
  \bibinfo{editor}{Wolfgang~J. Paul}, editors: {\sl
  \bibinfo{booktitle}{CHARME}}, {\sl \bibinfo{series}{Lecture Notes in Computer
  Science}} \bibinfo{volume}{3725}. \bibinfo{publisher}{Springer-Verlag}, pp.
  \bibinfo{pages}{162--175}.
\bibitemend

\bibitemstart{McMillan92THsmv}
\bibinfo{author}{Ken~L. McMillan} (\bibinfo{year}{1992}):
  \emph{\bibinfo{title}{Symbolic model checking - an approach to the state
  explosion problem}}.
\newblock \bibinfo{type}{Ph.D. thesis}, \bibinfo{school}{Carnegie Mellon
  University}.
\bibitemend

\bibitemstart{DBLP:conf/lics/VardiW86}
\bibinfo{author}{Moshe~Y. Vardi} \& \bibinfo{author}{Pierre Wolper}
  (\bibinfo{year}{1986}): \emph{\bibinfo{title}{An Automata-Theoretic Approach
  to Automatic Program Verification (Preliminary Report)}}.
\newblock In: {\sl \bibinfo{booktitle}{LICS}}. \bibinfo{publisher}{IEEE
  Computer Society}, pp. \bibinfo{pages}{332--344}.
\bibitemend

\bibitemstart{VBBBipdps09divine}
\bibinfo{author}{Kees Verstoep}, \bibinfo{author}{Henri~E. Bal},
  \bibinfo{author}{Jiri Barnat} \& \bibinfo{author}{Lubos Brim}
  (\bibinfo{year}{2009}): \emph{\bibinfo{title}{Efficient large-scale model
  checking}}.
\newblock In: {\sl \bibinfo{booktitle}{IPDPS}}. \bibinfo{publisher}{IEEE}, pp.
  \bibinfo{pages}{1--12}.
\bibitemend

\bibitemstart{Hao09EEDM}
\bibinfo{author}{Hao Wang} \& \bibinfo{author}{Wendy MacCaull}
  (\bibinfo{year}{2009}): \emph{\bibinfo{title}{An Efficient Explicit-time
  Description Method for Timed Model Checking}}.
\newblock In: {\sl \bibinfo{booktitle}{Parallel and Distributed Methods in
  verifiCation, 8th International Workshop, PDMC 2009, Held as Part of the
  Formal Methods Week 2009, Eindhoven, the Netherlands, November 2-6, 2009}}.
\bibitemend

\bibliographyend
\end{thebibliography}

\end{document}